\documentclass[12pt]{iopart}
\usepackage{setstack}
\usepackage{amssymb}
\usepackage{bm}
\usepackage{graphicx}
\begin{document}
%--------------------------------------------------------------------
%--------------------------------------------------------------------
%-----------------------------------------------------------------
\title{Stable gravastars --- an alternative to black holes?}
%-----------------------------------------------------------------
%-----------------------------------------------------------------
\author{Matt Visser\;\dag\;
\footnote[2]{\sf http://www.mcs.vuw.ac.nz/\~{}visser} 
and David L. Wiltshire\;\P\;
\footnote[4]{\sf http://www.phys.canterbury.ac.nz/\~{}dlw24}}
%-----------------------------------------------------------------
\address{\dag\; School of Mathematical and Computing Sciences, 
Victoria University of Wellington, New Zealand}
%-----------------------------------------------------------------
\address{\P\; Department of Physics and Astronomy, University of Canterbury, 
Private Bag 4800, Christchurch, New Zealand}
%-----------------------------------------------------------------
\eads{\mailto{matt.visser@vuw.ac.nz},
\mailto{d.wiltshire@phys.canterbury.ac.nz}}
%-----------------------------------------------------------------
%-----------------------------------------------------------------
%-----------------------------------------------------------------
\begin{abstract}
%-----------------------------------------------------------------
  The ``gravastar'' picture developed by Mazur and Mottola is one of a
  very small number of serious challenges to our usual conception of a
  ``black hole''. In the gravastar picture there is effectively a
  phase transition at/ near where the event horizon would have been
  expected to form, and the interior of what would have been the black
  hole is replaced by a segment of de Sitter space. While Mazur and
  Mottola were able to argue for the thermodynamic stability of their
  configuration, the question of dynamic stability against spherically
  symmetric perturbations of the matter or gravity fields remains
  somewhat obscure. In this article we construct a model that shares
  the key features of the Mazur--Mottola scenario, and which is
  sufficiently simple for a full dynamical analysis. We find that
  there are \emph{some} physically reasonable equations of state for
  the transition layer that lead to stability.

\vskip 0.50cm
\noindent
  Dated:  23 October 2003; \LaTeX-ed \today
\\
arXiv: gr-qc/0310107

%-----------------------------------------------------------------
\end{abstract}
%-----------------------------------------------------------------

%-----------------------------------------------------------------
\centerline{To appear in {\bf Classical and Quantum Gravity}}
%-----------------------------------------------------------------
%-----------------------------------------------------------------

%-----------------------------------------------------------------
\maketitle
%-----------------------------------------------------------------

%-----------------------------------------------------------------
% LOCAL DEFINES
%-----------------------------------------------------------------
\def\d{{\mathrm{d}}}
\def\be{\begin{equation}}
\def\ee{\end{equation}}
\def\implies{\Rightarrow}
\def\lsim{\mathop{\hbox{${\lower3.8pt\hbox{$<$}}\atop{\raise0.2pt\hbox{$\sim$}}
$}}} \def\gsim{\mathop{\hbox{${\lower3.8pt\hbox{$>$}}\atop{\raise0.2pt\hbox{$
\sim$}}$}}} \def\goesas{\mathop{\sim}\limits}
\def\w#1{\;\hbox{#1}\;}  
\def\pt{\partial} 
\def\vth{\vartheta}
\def\al{\alpha} 
\def\la{\lambda}
%Preprint style line spacing
%\baselineskip=15pt plus.2pt minus.1pt

\def\ns#1{_{\mathrm{#1}}}

%-----------------------------------------------------------------
\section{Introduction}
%-----------------------------------------------------------------

Whereas researchers in the relativity community (and the bulk of the
astrophysics and particle physics communities) are by and large happy
with our understanding of classical black holes, there is a certain
amount of polite dissent. Such dissent ranges from a careful sceptical
analysis of what it physically means to observe an event
horizon~\cite{marek}, through alternative models for compact
objects~\cite{boson}, to more radical proposals that drastically
modify the physics in the region where the event horizon would
otherwise be expected to
form~\cite{gerard,berezin,laughlin,laughlin2,laughlin3,gravastar}.

In particular, the ``gravastar'' ({\it gra}vitational {\it va}cuum
{\it star}) picture recently developed by Mazur and
Mottola~\cite{gravastar} is one of a very small number of serious
challenges to our usual conception of a ``black hole''. In the
gravastar picture there is effectively a phase transition at or near
the location where the event horizon would have been expected to form.
The interior of what would have been the black hole is replaced by a
suitably chosen segment of de Sitter space. (See also Laughlin
\emph{et al.}~\cite{laughlin,laughlin2,laughlin3} for a
similar proposal.) While Mazur and Mottola were able to make
considerable progress with their proposal, and to argue for the
thermodynamic stability of their configuration, the question of
whether their model enjoys full dynamic stability against spherically
symmetric perturbations of the matter and/ or gravity fields remains
an open question. In this article we construct a simplified model that
shares the key features of the Mazur--Mottola scenario, and which is
sufficiently simple to be amenable to a full dynamical analysis. We
find that there are \emph{some} physical equations of state for the
transition layer that lead to stability.

%------------------------------------------------------------
\section{The physical model}
%------------------------------------------------------------

To set the stage, we note that the Mottola-Mazur gravastar is an
onion-like construction which has five layers (including two thin
shells):
\begin{itemize}
\item
An external Schwarzschild vacuum, with energy density, $\rho=0$, and
pressure, $p=0$.
\item
A thin shell, with surface density $\sigma_+$ and surface tension
$\vartheta_+$; with radius $r_+\gsim2M$.
\item
A (relatively thin) finite-thickness shell of stiff matter with equation
of state $p=\rho$; straddling $r=2M$ where the horizon would in normal
circumstances have formed.
\item
A second thin shell; with radius $r_-\lesssim2M$, and with surface
density $\sigma_-$ and surface tension $\vartheta_-$.
\item
A de Sitter interior, with $p=-\rho$.
\end{itemize}
The two thin shells are used to ``confine'' the stiff matter in a
transition layer straddling $r=2M$, while the energy density in the de
Sitter vacuum is chosen to satisfy
\begin{equation}
{4\pi\over3} \rho (2M)^3 = M,
\end{equation} 
so that, in the approximation where the transition layer is neglected,
\emph{all} of the mass of the resulting object can be traced back to
the energy density of the de Sitter vacuum.

In order to dynamically study the gravitational stability of this
setup, we will simplify the model even further: We shall combine the
thick shell of stiff matter, and the two thin shells, into a single
thin shell. This reduces the gravastar to a simple three-layer model:
\begin{itemize}
\item
An external Schwarzschild vacuum, $\rho=0=p$.
\item
A single thin shell, with surface density $\sigma$ and surface
tension $\vth$; with radius $a\gsim2M$.
\item
A de Sitter interior, $p=-\rho$.
\end{itemize}
To avoid forming an event horizon, we shall demand
\begin{equation} 
{4\pi\over3} \rho (2M)^3 \lesssim M. 
\end{equation} 
The closer we are to saturating this bound, the closer we will be to
the Mazur--Mottola scenario (and the closer we will be to attributing
all the mass of the gravastar to the energy of the de Sitter vacuum).

We will analyse the dynamical stability of this simplified three-layer
model. Initially, we had hoped to either unambiguously determine
stability or unambiguously determine instability (which would then
have killed the gravastar scenario). The actual results are much more
ambiguous: \emph{Some} physically reasonable equations of state for
the thin shell lead to stability, but that situation does not appear
to be completely generic --- some fine tuning seems to be necessary.

%------------------------------------------------------------
\section{The mathematical model}
%------------------------------------------------------------

Let us consider the class of geometries
\begin{equation}
\d s^2 = - \left[1-{2m(r)\over r}\right] \;\d t^2 + 
{\d r^2\over1-2m(r)/r} + r^2 (\d\theta^2 + \sin^2\theta\;\d\phi^2).
\label{E:metric}
\end{equation}
While less general than the class of all static spherically symmetric
geometries, this restricted class of metrics is more than sufficient
for our current needs, and includes both the Schwarzschild and the de
Sitter geometries. We will assume that two geometries of this type are
connected along a timelike hypersurface at $r=a(t)$, with spacelike
normal, $n^a$. By considering a point at fixed $\theta$ and $\phi$ and
making use of the definition
\begin{eqnarray}
-\d\tau^2 &=& 
-\;\left[1-{2m(a(t))\over a(t)}\right]\; {\d t}^2
+\;{1\over1-2m(a(t))/a(t)} \left[{\d a(t)\over \d t}\right]^2 \d t^2,
\label{tau}
\end{eqnarray}
we can reparameterize the position of the timelike hypersurface in
terms of $\tau$ --- the proper time along this hypersurface --- and so
determine $a(\tau)$. To understand the dynamics of the hypersurface we
adopt the Israel--Lanczos--Sen thin-shell formalism~\cite{shell}. The
induced metric on the shell,
\begin{equation}
h_{ab}=g_{ab}-n_a n_b,
\end{equation}
is given by
\begin{equation}
h_{ab} \; \d x^a\,\d x^b
= - \d\tau^2 + a(\tau)^2 \; (\d\theta^2 + \sin^2\theta\;\d\phi^2).
\end{equation}
We are interested in the extrinsic curvature
\begin{equation}
K_{ab} = {h_a}^c \, {h_b}^d \, \nabla_c n_d
\label{K:gen}
\end{equation}
so that we can then apply the junction conditions to relate the
discontinuity in extrinsic curvature to the surface stress-energy,
$S_{ab}$, located on the shell:
\begin{equation}
[[ K_{ab} ]] = -8\pi \left[ S_{ab} - {1\over2} S \; h_{ab} \right];
\qquad
[[ K_{ab} - K\; h_{ab} ]] = -8\pi \; S_{ab}.
\end{equation}
Here $[[X]]$ denotes the discontinuity in $X$ across the shell. A
modern exposition of the thin-shell formalism (in a rather different
physical context) can be found in~\cite{visser} to which we make
extensive reference in the interests of simplifying the current
presentation.

%------------------------------------------------------------
\subsection{Static shell}
%------------------------------------------------------------

For simplicity, let us first assume the shell is static. The extrinsic
curvature may be computed directly from the definition (\ref{K:gen}),
or alternatively in terms of Gaussian normal coordinates (GNC). Define
$n^a=\left({\pt/\pt\eta}\right)^a$, where
\begin{equation}
\d\eta = {\d r\over\sqrt{1-2m(r)/r}}\,.
\end{equation}
In GNC the extrinsic curvature is
\begin{equation}
K_{ab} = {1\over2} {\partial g_{ab}\over\partial r}\; 
{\partial r\over\partial\eta}\,,
\label{K:GNC}
\end{equation}
or in an orthonormal basis,
\begin{equation}
K_{\hat a\hat b} =\sqrt{1-2m(a)/a}\; \w{diag}
\left({\{m(a)/a\}'\over 1-2m(a)/a}\;,\,{1\over a}\;,\,{1\over a}\right)\,.
\end{equation}
Then in an orthonormal frame the surface stress-energy tensor is given
by $S_{\hat a\hat b}=\w{diag}\!(\sigma,-\vth,-\vth)$, and standard
manipulations yield both
\begin{equation}
\left[\left[ \sqrt{1-2m(a)/a} \;\; a^{-1} \right]\right] 
= -4\pi\sigma,
\label{E1:static}
\end{equation}
and
\begin{equation}
\left[ \left[ 
{ 1-m(a)/a - m'(a) \over a \; \sqrt{1-2m(a)/a} } 
\right]\right] = -8\pi\vth.
\label{E2:static}
\end{equation}
Thus there are only two algebraically independent components of the
extrinsic curvature and they can be related to surface density and
surface tension. (Compare, for instance, with~\cite{visser}, pp 179
ff.~\footnote{Note that equation (15.39b) on page 179 of~\cite{visser}
  has a typographical error and should read
\begin{eqnarray*}
\vth = -{1\over8\pi} \cdot 
\left\{ 
- \kappa^{\hat\tau\hat\tau}+ \kappa^{\hat\theta\hat\theta}
\right\}.
\end{eqnarray*}
} For calculationally similar though physically distinct
presentations, see
also~\cite{surgery,brane-surgery,edge,brady,poisson}.)

Equation (\ref{E1:static}) may be recast as
\begin{equation}
 \sqrt{1-2m_+/a} = \sqrt{1-2m_-/a}\; -4\pi\sigma\; a\,,
\label{junction1}\end{equation}
where we use the notation $m_+\equiv m(a_+)$ and $m_-\equiv m(a_-)$.
If we square both sides of equation (\ref{junction1}) and rearrange it
follows that
\begin{equation}
 8 \pi\sigma\; a^2 \sqrt{1-2m_-/a} = 16\pi^2 \sigma^2 a^3 +2(m_+-m_-).
\end{equation}
A further manipulation of this sort then yields
\begin{equation}
16\pi^2 \; \sigma^2 a^4 +4m_+m_- = 
\left[8\pi^2 \sigma^2 a^3 + (m_++m_-) \right]^2.
\label{junction2}\end{equation}
If we wish to solve (\ref{junction2}) for $a$ as a function of the
other parameters, we are dealing with a sextic. In contrast, solving
for any of one of $m_+$, $m_-$, or $\sigma$ is no worse than a simple
quadratic. One can simplify equation (\ref{junction2}) by expressing
it in terms of $m_s = 4\pi\,\sigma\,a^2$, which represents the mass of
the thin shell itself. We then obtain a [static] ``master equation'',
which directly relates the three masses ($m_+$, $m_-$, and $m_s$) to
the radial location of the shell, $a$:
\begin{equation}
m_s^2 + 4 m_+ m_- = \left[m_s^2/(2a)+ (m_+ + m_-) \right]^2.
\label{master_static}
\end{equation}
We have gone through this analysis in some detail as we will want
to re-use and generalize this calculation in the dynamic case.

The master equation (\ref{master_static}) is only one single equation;
one would simultaneously have to solve the surface tension equation.
In many ways the best approach for the static shell is prescriptive:
choose the interior and exterior geometries arbitrarily, then choose
$a$, and then calculate the shell energy density $\sigma$ and surface
tension $\vth$ from equations (\ref{E1:static}) and (\ref{E2:static}).
Of course, this analysis presently gives no information about
stability; we are simply \emph{assuming} a static shell. Developing a
dynamical stability analysis is our next concern.

%------------------------------------------------------------
\subsection{Dynamic shell}
%------------------------------------------------------------

We now add time dependence by allowing the shell to move radially, so
that the proper time, $\tau$, at points on the shell is given by
(\ref{tau}). It follows from (\ref{tau}) that the (unit) 4-velocity of
a point on the hypersurface (at fixed $\theta$ and $\phi$) is
\begin{equation}
V^a = 
\left( {\sqrt{\Delta+\dot a^2}\over\Delta},
\dot a, 0, 0 \right),
\label{V:def}
\end{equation}
where $\Delta\equiv1-2m(a)/a$, $\dot a = da/d\tau$ and we have chosen
$\tau$ to be future-pointing with respect to $t$. The unit normal to
the shell is obtained from (\ref{V:def}) using the orthogonality
condition, $n_a V^a=0$, giving
\begin{equation}
n^a = \left( {\dot a\over\Delta} , 
\sqrt{\Delta+\dot a^2} ,0,0\right).
\label{n:def}
\end{equation}
There is an overall sign ambiguity in defining the components of the
unit normal (\ref{V:def}), which is fixed by demanding that the chart
of the Gaussian normal coordinates $(\tau,\eta)$ be consistently
oriented with respect to the chart of local coordinates $(t,r)$, the
Jacobian of the transformation being
\begin{equation}
{\pt(\tau,\eta)\over\pt(t,r)}=\left(\begin{array}{ccc}
\sqrt{\Delta+\dot a^2} & -\Delta^{-1}\dot a\\
-\dot a & \Delta^{-1}\sqrt{\Delta+\dot a^2}\end{array} \right)\,,
\end{equation}
with unit determinant and inverse
\begin{equation}
{\pt(t,r)\over\pt(\tau,\eta)}=\left(\begin{array}{ccc}
\Delta^{-1}\sqrt{\Delta+\dot a^2} & \Delta^{-1}\dot a\\
\dot a & \sqrt{\Delta+\dot a^2}\end{array} \right)\,.
\label{jac2}
\end{equation}

The components of the extrinsic curvature in GNC may be computed using
(\ref{K:GNC}) and (\ref{jac2}), or alternatively directly from the
definition (\ref{K:gen}). In fact, the second approach is more direct
in the present case. On account of the orthogonality relation,
combined with the definition of the induced metric, we have
${h_a}^b\,V^a= V^b$, so that it immediately follows that
\begin{eqnarray}
K_{\hat\tau\hat\tau}=K_{\tau\tau}&=&V^a\,V^b\,K_{ab}
\nonumber\\
&=&V^a\,V^b\,\nabla_an_b\nonumber\\
&=&-n_b\,A^b\,,
\end{eqnarray}
where $A^b\equiv V^a\nabla_aV^b$ is the 4-acceleration of a point on
the shell. Since the 4-acceleration and 4-velocity are orthogonal we
must have $A^b=An^b$, so that
\begin{equation}
K_{\hat\tau\hat\tau}=-A
\end{equation}
where $A$ is the magnitude of the 4-acceleration of the shell. The
angular components of the extrinsic curvature are easily found to be
\begin{equation}
K_{\hat\theta\hat\theta}=K_{\hat\phi\hat\phi}={1\over a}\sqrt{\Delta+
\dot a^2}\,.
\end{equation}

Imposing the junction conditions, similarly to the case of the static
shell, we then have
\begin{equation}
\sigma = -{1\over4\pi a} 
\left[\left[\sqrt{1-2m(a)/a +\dot a^2}\right]\right], 
\label{E1:gen}
\end{equation}
and
\begin{equation}
\vth= -{1\over8\pi} 
\left[\left[ {\sqrt{1-2m(a)/a+\dot a^2}\over a} + A \right] \right],
\label{E2:a}
\end{equation}
Since the restricted class of geometries we are considering possess a
timelike Killing vector, $k^a=\left({\pt/\pt t}\right)^a$, the
4-acceleration is easily calculated. Following reference~\cite{visser},
p.~183, we can use both
\begin{equation}
{d\over d\tau} (k_a V^a) = - A \dot a,
\end{equation}
and
\begin{equation}
{d\over d\tau} (k_a V^a) = 
-{d\over d\tau} 
\left\{ {\sqrt{1-2m/a+\dot a^2}}\right\},
\end{equation}
to give the simple relation
\begin{equation}
A = 
{\ddot a - \{m/a\}' \over \sqrt{1-2m/a+\dot a^2}}.
\label{A:gen}
\end{equation}
This gives the 4-acceleration of the shell in terms of a combination
of kinematic ($\ddot a$, $\dot a$) and gravitational ($m(a)/a$)
properties.
%Compare with~\cite{visser} equation (15.62).
Substituting (\ref{A:gen}) in (\ref{E2:a}) we then have
\begin{equation}
\vth = -{1\over8\pi a} 
\left[\left[{1-m/a - m' +\dot a^2 + a \ddot a
\over\sqrt{1-2m/a+\dot a^2} }
\right]\right].
\label{E2:gen}
\end{equation}
We now have \emph{explicit} formulae for both surface density
(\ref{E1:gen}) and surface tension (\ref{E2:gen}).  These equations
are the natural generalization of equations (15.63) and (15.64) of
reference~\cite{visser}, where a somewhat more restricted class of
metric was considered in a different context.

%------------------------------------------------------------
\subsection{Conservation laws}
%------------------------------------------------------------

Equation (\ref{E2:gen}) for $\vth$, though manageable, is still a
little messy. There is no need to use it directly, however, since
equations (\ref{E1:gen}) and (\ref{E2:gen}) are related on account of
energy-momentum conservation. It is easily verified by straightforward
manipulations of (\ref{E1:gen}) and (\ref{E2:gen}) that
\begin{equation}
{d\over d\tau} (\sigma a^2) = \vth {d\over d\tau} (a^2).
\label{cons_law}
\end{equation}

We note that in general one would expect the the r.h.s.\ of
(\ref{cons_law}) to be supplemented by a term of the form
\begin{equation}
\hbox{flux term}=\left[\left[ T_{ab} V^a n^b \right]\right]\,,
\end{equation}
corresponding to the net discontinuity in the momentum flux, $ F_a =
T_{ab} \; V^b$, impinging on the shell. Such a term is absent in the
case of the special class of metrics (\ref{E:metric}) considered here,
for which
\begin{equation} 
\rho = -p_r = {1\over4\pi} {m'\over r^2},
\end{equation} 
and
\begin{equation} 
p_t = -{1\over8\pi}{m''\over r}.
\end{equation} 
Since the density equals minus the radial pressure, when restricted to
the $r$--$t$ plane $T_{ab}\propto g_{ab}$. But since both $V$ and $n$
lie in the $r$--$t$ plane this implies
\begin{equation} 
T_{ab} V^a n^b \propto g_{ab} V^a n^b = 0, 
\end{equation}
and the flux terms on both sides of the shell automatically vanish.
This is a special feature of the ``sufficiently general'' class of
geometries (\ref{E:metric}). By contrast, in the general spherically
symmetric geometry this flux term would be somewhat difficult to deal
with.

Consequently, if we are provided with an equation of state for the
shell $\sigma = \sigma(\vth)$, [or $\vth=\vth(\sigma)$], we can
formally integrate the conservation equation (\ref{cons_law}) to
determine $\sigma(a)$ as a function of $a$. Once we have that
function, \emph{all} the dynamics will be encoded in the single
dynamical equation (\ref{E1:gen}), the result being true subject only
to our ``general enough'' restriction on the metric (\ref{E:metric}).
\bigskip

%------------------------------------------------------------
\section{The master equation}
%------------------------------------------------------------

%------------------------------------------------------------
\subsection{Derivation}
%------------------------------------------------------------

Let us re-write the dynamical $\sigma$ equation (\ref{E1:gen}) as
\begin{equation}
\left[ \left[ \sqrt{1-2m(a)/a+\dot a^2} \right] \right] 
= -4\pi\sigma(a)\; a,
\label{E:master}
\end{equation}
where we are now keeping arbitrary time dependence in the form of
$\dot a$. That is
\begin{equation}
 \sqrt{1-2m_+(a)/a+\dot a^2} = \sqrt{1-2m_-(a)/a+\dot a^2} -4\pi\sigma(a)\; a.
\end{equation}
A series of steps identical to those that lead from (\ref{junction1})
to (\ref{master_static}) now yields
\begin{equation}
{m_s}^2(1+\dot a^2) + 4 m_+ m_- = \left[{m_s^2\over2a}+(m_++m_-)\right]^2\,,
\label{master_dynamic}
\end{equation}
where, as before $m_s = 4\pi \sigma a^2$ is the mass of the thin
shell.  This dynamic ``master equation'' is of the form of an ``energy
equation'' for a nonrelativistic particle,
\begin{equation}
{1\over2} \dot a^2 + V(a) = E,
\label{E:energy}
\end{equation}
with ``potential''
\begin{equation}
\label{E:potential}
V(a) = {1\over2}
\left\{1 + {4 m_+(a) m_-(a) \over m_s^2(a)} -
\left[{m_s(a)\over 2a} + {(m_+(a)+m_-(a))\over m_s(a)} \right]^2\right\},
\end{equation}
and ``energy'' $E=0$. 

There will then be a strictly stable solution for the shell (stable
against spherically symmetric radial oscillations) if and only if
there is \emph{some} $m_s(a)$ and \emph{some} $a_0$ such that we
simultaneously have
\begin{equation}
V(a_0) = 0; \qquad\quad V'(a_0)=0; \qquad\quad V''(a_0) > 0.
\label{V_cond}
\end{equation}
A small quirk in this relativistic calculation is that since
$E\equiv0$, the situation where $V(a)\equiv0$, which in
nonrelativistic mechanics corresponds to neutral equilibrium, is now
converted to a situation of stable equilibrium (since now, because one
is not free to increase the ``energy'' $E$, one has $\dot a \equiv
0$).

There is a less stringent notion of stability that is also useful,
that of ``bounded excursion''. Suppose we have $a_2>a_1$ such that
\begin{equation}
V(a_1)=0; \quad  V'(a_1)\leq 0; \quad V(a_2)=0; \quad  V'(a_2)\geq 0;
\end{equation}
with $V(a) < 0$ for $a\in(a_1,a_2)$. In this situation the motion of
the shell remains bounded by the interval $(a_1,a_2)$. Although not
strictly stable, since the shell does in fact move, this notion of
``bounded excursion'' more accurately reflects some of the aspects of
stability naturally arising in nonrelativistic mechanics. In
particular, it is simply a version of the standard stability criterion
for orbits about the fixed point that would exist if we were free to
arbitrarily specify the constant $E$ on the r.h.s.\ of
(\ref{E:energy}), and corresponds to orbits about the fixed point
which are ``stable but not asymptotically stable''.  In the present
context, perturbing the potential by adding a small negative offset
\begin{equation}
V(a) \to V(a) - \epsilon^2
\end{equation}
will generically convert a strictly stable potential to one exhibiting
``bounded excursion''.

%------------------------------------------------------------
\subsection{Inverting the potential}
%------------------------------------------------------------

Suppose we now take the potential $V(a)$, and the masses $m_\pm(a)$ as
given. Then equation (\ref{E:potential}) yields a quadratic equation
for $m_s^2(a)$:
\begin{eqnarray}
m_s^2(a) &=& 
2a^2 \Bigg\{
1- {2 V(a)} - {m_+(a)\over a} - {m_-(a)\over a} 
\nonumber\\
&&
\pm 
\sqrt{1- {2V(a)}- {2m_+(a)\over a} } \; 
\sqrt{1-{2V(a)}- {2m_-(a)\over a}} 
\Bigg\}. 
\end{eqnarray} 
It is important to note that we have not chosen any specific equation
of state for the shell; we have only used the assumption that
\emph{some} equation of state exists. As always when solving a
quadratic, there is a risk that one of the roots is unphysical. In
this case the ($+$) sign is unphysical and the ($-$) sign the
physically interesting branch. This may most easily be verified by
setting $m_+(a)=m_-(a)$ in which case the spacetime geometry is
completely continuous across the location of the shell, so that its
mass $m_s$ (and surface energy density $\sigma$ and surface tension
$\vth$) must be zero. Therefore the surface energy density $\sigma(a)$
is given by the expression
\begin{eqnarray}
\fl
\sigma(a)=
\nonumber
\\
\fl \pm {\sqrt2\over 4\pi a}
\sqrt{ 1 - {2 V(a)}- {m_+(a)\over a} - {m_-(a)\over a} - 
\sqrt{1- {2 V(a)}-{2m_+(a)\over a}} 
\sqrt{1- {2 V(a)}-{2m_-(a)\over a}} }.
\nonumber
\\
\end{eqnarray}
We can now, by noting that the contents of the outer radical sign
above are a perfect square, extract the square root. The overall sign
is determined by the unique choice which leads to the above expression
being the compatibility equation between equations (\ref{E:energy})
and (\ref{E:master}).  This now yields:
\begin{eqnarray}
\sigma(a) &=& -{1\over 4\pi a}
\left\{ 
\sqrt{1- {2 V(a)}-{2m_+(a)\over a}} - 
\sqrt{1- {2 V(a)}-{2m_-(a)\over a}} \right\}
\\
&=&
-{1\over 4\pi a}
\left[\left[
\sqrt{1- {2 V(a)}-{2m(a)\over a}}\;
\right]\right].
\label{E:eos}
\end{eqnarray}

There is a subtlety one should be aware of when comparing equation
(\ref{E:master}) with (\ref{E:eos}). Whereas in (\ref{E:master}) we
are simply determining the energy density on a shell if it happens to
have ``velocity'' $\dot a$ at radius $a$, here in (\ref{E:eos}) we are
determining something considerably more powerful based on a fully
dynamic analysis, viz., what the surface energy density would have to
be as a function of $a$ in order to be compatible with the specified
potential $V(a)$. Proceeding further, from the conservation equation
\begin{equation}
\vth = \sigma+{1\over2} \, a \, {\d\sigma\over \d a},
\end{equation}
we can calculate $\vth(a)$. Analytically, the result is 
\begin{equation}
\vth(a) = -{1\over 8\pi a} 
\left[ \left[ 
{
1-{2 V(a)} - {m(a)/a} - {a V'(a)} - m'(a) 
\over
\sqrt{1-2 V(a)-2m(a)/a}
}
\right]\right].
\label{E:tension}
\end{equation}
Recall that $m_\pm(a)$ and $V(a)$ are at this stage arbitrary
functions that we are still free to prescribe. These expressions for
$\sigma(a)$ and $\vth(a)$ --- equations (\ref{E:eos}) and
(\ref{E:tension}) --- now determine the equation of state
parametrically. In summary
\begin{itemize}
\item
\emph{Either} we choose an equation of state in the form of a
specific $\sigma(\vth)$ and integrate the conservation equation to
determine $\sigma(a)$ [and hence $m_s(a)$ and hence $V(a)$];
\item
\emph{or} we choose a potential in the form $V(a)$ and parametrically
extract the corresponding equation of state using the algorithm just
presented.
\end{itemize}

%------------------------------------------------------------
\subsection{Specialization of gravastar geometry}
%------------------------------------------------------------

\noindent
We will now specialize to a specific model for the gravastar, which
simplifies the Mazur--Mottola model to a thin shell separating two bulk 
regions:
\begin{itemize}
\item
Outside, a Schwarzschild region $T_{ab}=0$.
\item
Inside, a de Sitter region $T_{ab}\propto g_{ab}$.
\end{itemize}
Thus $m_+$ and $m_-$ are no longer arbitrary. In view of our
assumptions we can simply write $m_+=M$, the total external mass of
the system, while
\begin{equation}
m_- = (4\pi/3) \rho a^3 = k a^3,
\end{equation}
where $\rho$ is the energy density in the de Sitter vacuum and we have
introduced the parameter $k$ simply to minimize the explicit
appearance of factors of $4\pi/3$. Keeping $V(a)$ arbitrary, our
previous analysis implies
\begin{equation}
\sigma(a) \equiv {1\over 4\pi a} 
\left\{ \sqrt{1 -2V(a)- 2 k a^2} - \sqrt{1-2V(a)-{2M\over a}}
\right\},
\label{E:sigma-V}
\end{equation}
while for $\vth(a)$ we have
\begin{eqnarray}
\fl
\vth(a)&\equiv& {1\over8\pi a} 
\Bigg\{
{1-2V(a)-a\;V'(a)-4ka^2
\over
\sqrt{1 -2V(a)- 2 k a^2} }
-
{1-2V(a)-a\;V'(a)-M/a
\over
\sqrt{1-2V(a)-{2M/ a}} }
\Bigg\}.
\label{E:theta-V}
\end{eqnarray}

These now are a pair of parametric equations determining the equation
of state for any 3-layer Mazur--Mottola model with exterior
Schwarzschild geometry and interior de Sitter geometry in terms of a
freely specified $V(a)$. (For formally similar analyses in physically
different contexts, see~\cite{brady,poisson}.) If we want the system
to be strictly stable at some $a_0$ then we need to enforce
equation (\ref{V_cond}).
There are two obvious examples of such behaviour:
\begin{itemize}
\item
$V(a)\equiv 0$, a degenerate, but physically important, case
corresponding to $\dot a\equiv 0$.
\item $V(a) = {1\over2}(a-a_0)^2\; f(a)$, where $f(a)$ is an arbitrary
  positive function which is regular at $a_0$. In this situation the
  stability is in a sense trivial: The master equation has a unique
  solution at $a=a_0$ and $\dot a=0$, and all possibility of motion is
  excluded by fiat.
\end{itemize}
More generally we could consider cases of ``bounded excursion'' by
choosing $V(a) = {1\over2}(a-a_0)^2\; f(a) - \epsilon^2$. For
sufficiently small $\epsilon$ there will be distinct roots $a_1$ and
$a_2$ where $V(a)=0$, with $a_1<a_0<a_2$. The motion of the shell will
be of ``bounded excursion'' and forced to remain in the interval
$(a_1,a_2)$.

These observations are now enough to guarantee that there are large
classes of potential $V(a)$ --- and so large classes of equation of
state $\sigma(\vartheta)$ --- for which the system is stable against
radial oscillations. Of course there are also large classes of
potential for which the system is unstable. Depending on one's views
regarding the gravastar model this may be interpreted as either good
news or bad news.

%------------------------------------------------------------
\subsection{\label{secV} The $V(a)\equiv 0$ case}
%------------------------------------------------------------

We will now investigate the $V(a)\equiv 0$ case. This is an important
case as it corresponds to a completely static gravastar with $\dot
a\equiv0$. Furthermore, since equations (\ref{E:sigma-V}) and
(\ref{E:theta-V}) have no explicit dependence on $V''(a)$ or higher
derivatives of $V$, the analysis we will present suffices to determine
the relationship between the energy density, $\sigma$, and surface
tension, $\vth$, of the shell at any specific $a_0$ for which
$V(a_0)=0$ and $V'(a_0)=0$.

Setting $V(a)\equiv 0$ in (\ref{E:sigma-V}) and (\ref{E:theta-V}) we find
\begin{equation}
\sigma(a) \equiv {1\over 4\pi a} 
\left\{ \sqrt{1 - 2 k a^2} - \sqrt{1-{2M\over a}}
\right\},
\label{E:sigma-0}
\end{equation}
and
\begin{equation}
\vth(a)\equiv {1\over8\pi a}
\left\{
{1-4ka^2
\over
\sqrt{1 - 2 k a^2} }
-
{1-M/a
\over
\sqrt{1-2M/ a}}
\right\}.
\label{E:theta-0}
\end{equation}
It is easy to check that the strong energy condition
\begin{equation}
\sigma - 2\vth > 0
\end{equation}
is satisfied for all values of the parameters for which real solutions
exist. The null energy condition is consequently also
satisfied. However, other energy conditions are possibly violated. (We
note that there are good reasons for not taking the energy conditions
too seriously as fundamental physics~\cite{twilight}. Nevertheless the
energy conditions are still useful as a crude measure of how peculiar
a certain matter distribution is.)

Since we assume $M>0$ in order that $\sigma$ and $\vth$ be real, we
require $a>2M$ and $a\leq1/\sqrt{2k}$. Consequently $k<1/(8M^2)$.
The surface energy function $\sigma(a)$ ranges from the positive value
\begin{equation}
\sigma(2M) = {\sqrt{1-8kM^2}\over8\pi M}\,,
\label{sigma_bound}
\end{equation}
at $a=2M$, to the negative value
\begin{equation}
\sigma\left({1\over\sqrt{2k}}\right) =
{-1\over4\pi}\sqrt{2k\left(1-2M\sqrt{2k}\right)}\,;
\end{equation}
at $a=1/\sqrt{2k}$, and crosses zero at $a=(M/k)^{1/3}$. Thus the weak
energy condition is satisfied for $2M<a< (M/k)^{1/3}$, but violated
for $(M/k)^{1/3}\le a<1/\sqrt{2k}$.

The status of the dominant energy condition, $|\vth|\leq \sigma$, is
more complicated and depends crucially on the value of the
dimensionless parameter $kM^2$. The surface tension $\vth$ itself is
negative on the whole interval $2M<a<1/\sqrt{2k}$, and diverges as
$\vth\to-\infty$ at the endpoints. To get the full equation of state
$\vth(\sigma)$ we can simply obtain a graphical solution from
(\ref{E:sigma-0}) and (\ref{E:theta-0}) by plotting $\theta(a)$ versus
$\sigma(a)$ parametrically.

As an example, we provide several plots exploring the case
$kM^2=1/18$, for which $a\ns{min}=2M$ and $a\ns{max} =
1/\sqrt{2k}=3M$.  In this case the dominant energy condition is
violated throughout the entire range $a\in(2M,3M)$.

%------------------------------------------------------------
\begin{figure}[htb]
\vbox{
\vskip 10 pt
\centerline{\scalebox{0.75}{\includegraphics{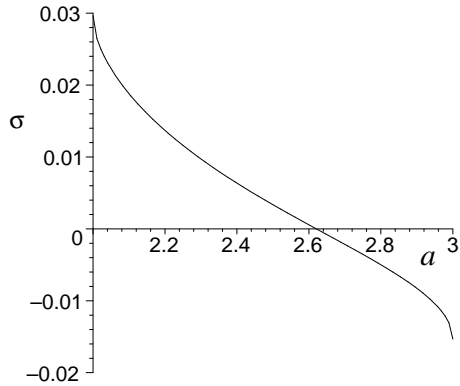}}}
\caption{%
{\sl Surface energy density, $\sigma$ (in units $M^{-1}$), as a function of
radius, $a$ (in units $M$). 
($kM^2=1/18$; $V(a)\equiv0$.)}}
\label{sigma_la18}} 
\end{figure}
%------------------------------------------------------------

%------------------------------------------------------------
\begin{figure}[htb]
\vbox{
\vskip 10 pt
\centerline{\scalebox{0.75}{\includegraphics{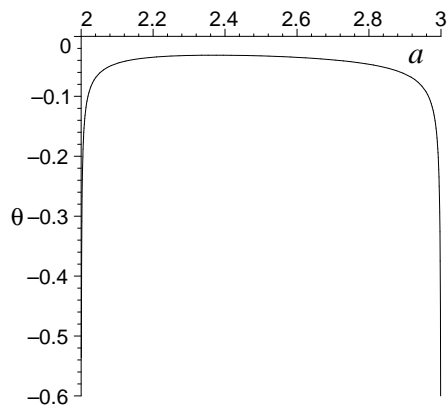}}}
\caption{%
{\sl Surface tension, $\vth$ (in units $M^{-1}$ ), as a function of radius,
$a$ (in units $M$). 
($kM^2=1/18$; $V(a)\equiv0$.)}}
\label{theta_la18}} 
\end{figure}
%------------------------------------------------------------

%------------------------------------------------------------
\begin{figure}[htb]
\vbox{
\vskip 10 pt
\centerline{\scalebox{0.75}{\includegraphics{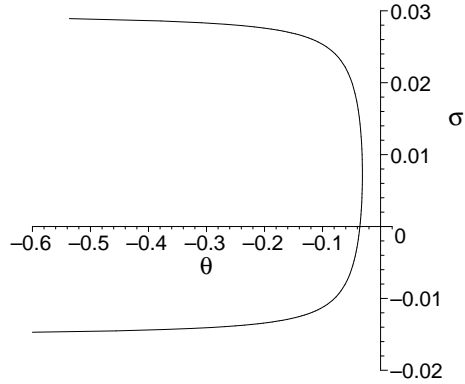}}}
\caption{%
 {\sl Equation of state: Surface energy density as a function of
 surface tension. ($kM^2=1/18$; $V(a)\equiv0$.)}}
\label{eos1}}
\end{figure}
%------------------------------------------------------------

%------------------------------------------------------------
\begin{figure}[htb]
\vbox{
\vskip 10 pt
\centerline{\scalebox{0.75}{\includegraphics{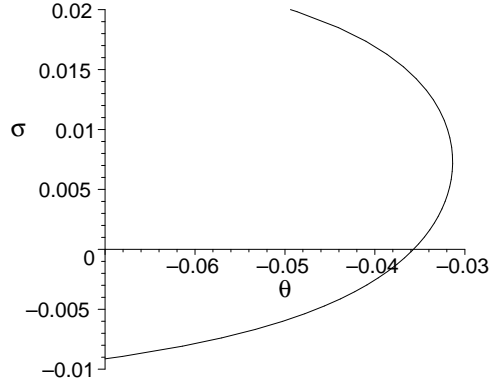}}}
\caption{%
 {\sl Equation of state: Enlargement of the central region of
 figure~\ref{eos1} --- surface energy density as a function of surface
 tension.  ($kM^2=1/18$; $V(a)\equiv0$.)}} }
\end{figure}
%------------------------------------------------------------

\begin{figure}[htb]
\vbox{
\vskip 10 pt
\centerline{\scalebox{0.75}{\includegraphics{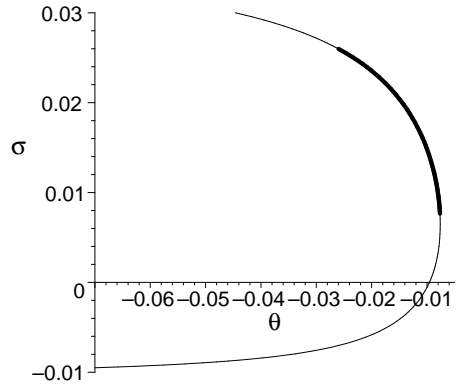}}}
\caption{%
 \label{la72}
 {\sl Equation of state: Enlargement of the central region for an example
 in which the dominant energy condition is satisfied. Surface energy density
 as a function of surface tension for the case $kM^2=1/72$; $V(a)\equiv0$.}
 Parameter values for which the dominant energy condition is violated are
 shown by a thin line, and parameter values for which the dominant energy
 condition is satisfied, viz., $2.124319\,M<a<3\,M$, are shown by a thick
 line.} }
\end{figure}

As a second specific example we plot the case $kM^2=1/72$, for which
$a\ns{max}=1/\sqrt{2k}=6M$, in figure~\ref{la72}.  In this case the
dominant energy condition is satisfied for $2.1243194234\;M \le
a\le3M$, and violated throughout the rest of the range
$a\in(2M,6M)$. The ``least stiff'' equation of state for this example
is obtained at $a=2.5764849609\,M$ when
$\vth=-0.6912774322\,\sigma$. We have simply shown the central region
of the equation of state $\sigma(\vth)$ in figure~\ref{la72}, since
the functions $\sigma(a)$ and $\vth(a)$ have the same qualitative form
as the case depicted in figures~\ref{sigma_la18} and~\ref{theta_la18},
the only difference being that $a$ now ranges from $a=2M$ to $a=6M$.

%------------------------------------------------------------
%------------------------------------------------------------

The range of parameter values for which it is possible to obtain
stable configurations which satisfy the dominant energy condition can
be quantified precisely. The critical case is that of a stiff matter
shell, for which $p=\rho$, or $\sigma=-\vth$. Applying this condition
to (\ref{E:sigma-V}) and (\ref{E:theta-V}) we have
\begin{equation}
\left(3-{5M\over a}\right)\sqrt{1-2ka^2}=
\left(3-8ka^2\right)\sqrt{1-{2M\over a}}.
\label{stiff-1}
\end{equation}
If we square both sides of (\ref{stiff-1}) and rearrange we obtain a
sextic in $a$,
\begin{eqnarray}
\fl
64k^2a^6-128k^2Ma^5-30ka^4+36kMa^3
+50kM^2a^2+12Ma-25M^2&=&0\,.
\label{stiff-2}
\end{eqnarray}
Only half of the roots of (\ref{stiff-2}) are potentially of interest
since (\ref{stiff-2}) also applies to the equation one would obtain by
changing the relative sign of the roots on each side of
(\ref{stiff-1}). Given that the l.h.s.\ of (\ref{stiff-1}) is positive
for $a>2M$, positivity of the r.h.s.\ of (\ref{stiff-1}) further
tightens the upper bound on $a$ to $a<\sqrt{3}/\sqrt{8k}$. There are
then at most two real solutions of (\ref{stiff-2}) in the range
$2M<a<\sqrt{3}/\sqrt{8k}$. (Note that in order for this range to be
non empty we must have $k M^2 < 3/32$, which is why the dominant
energy condition was always violated in our first example.)

To derive the condition for the existence the relevant thin shells it
is convenient to rewrite (\ref{stiff-2}) in terms of the dimensionless
parameters $\al=a/M$ and $\la=kM^2$, so that
\begin{equation}
64\la^2\al^6-128\la^2\al^5-30\la\al^4+36\la\al^3+50\la\al^2+12\al-25=0\,.
\label{stiff-3}
\end{equation}
Solutions of the required type exist if the discriminant of (\ref{stiff-3}),
when considered as a polynomial in $\al$, is negative, i.e.,
\begin{equation}
-14155776\,\la^7\left(200\,\la-27\right)^2 D_4(\la)<0\,,
\end{equation}
where
\begin{eqnarray}
\fl
D_4(\la)&\equiv&
\bigl(400000000\,\la^4-1054320000\,\la^3
+257041039\,\la^2-19516500\,\la+337500
\bigr)\,.
\label{quartic}
\end{eqnarray}
This means that the quartic $D_4(\la)$ must be positive.  Solving for
roots of this quartic numerically we find that
\begin{equation}
\la=kM^2\le\la\ns{cr}=0.0243045493773
\label{lambda_cr}
\end{equation}
is the critical condition for the existence of thin shells which
satisfy the dominant energy condition.

In the critical case $kM^2=\la\ns{cr}$ the dominant energy condition
is satisfied only for a stiff matter thin shell at $a=a\ns{cr}\equiv
2.3005600972496\, M$. For parameter values $0<kM^2<\la \ns{cr}$, there
will be a range of values $a_1<a<a_2$ over which the dominant energy
condition is satisfied. Figure~\ref{la72}\ shows one such example. The
bounds $a_{1,2}$ are given by the relevant solutions of
(\ref{stiff-2}), such that $2M<a_1<a\ns{cr}$ and
$a\ns{cr}<a_2<\sqrt{3}/\sqrt{8k}$. Thin shells with $a=a_{1,2}$ will
have stiff matter equations of state, while those with $a_1<a<a_2$
will have ``almost stiff'' equations of state.  Since the dominant
energy condition is the most stringent of the standard energy
conditions this construction implies that all the standard energy
conditions will be satisfied.

Although the construction used in this section has not yielded an
analytic expression for the equation of state in a form
$\sigma(\vartheta)$, it has nevertheless been ``effective'' --- this
construction does certainly demonstrate that there are physically
reasonable equations of state that lead to a stable gravastar
configuration for many values of the parameters $a$, $M$, and $k$.

It is also important to note that the $V(a)\equiv0$ examples
considered in this section are notable mainly for their relative
simplicity --- there is no physical reason for not choosing
$V(a)\neq0$, say $V(a) = {1\over2}(a-a_0)^2\; f(a) - \epsilon^2$, and
doing so opens new possibilities for stable shell configurations.

%------------------------------------------------------------
\section{Special gravastar geometries}
%------------------------------------------------------------
\subsection{$V\equiv 0$ equation of state --- Special limits}
%------------------------------------------------------------

\noindent
For a general equation of state of the thin shell,
two special limits of the $V\equiv0$ model are of interest:
\begin{itemize}
\item
$k\to 0$, corresponding to an empty interior.
\item
$k \to {1/(8M^2)}$, corresponding to the Mazur--Mottola limit.
\end{itemize}

%------------------------------------------------------------
%\subsubsection{Empty interior}
%------------------------------------------------------------

For an empty interior, results are straightforward, and can be
explicitly written as
\begin{equation}
\sigma(a) = {1-\sqrt{1-2M/a}\over4\pi a};
\end{equation}
\begin{equation}
\vth(a) = 
{1\over8\pi a} \; \left[1-{1-M/a\over\sqrt{1-2M/a}}\right]. 
\end{equation}
Though these formulae are very similar to standard results occurring
in the theory of thin shells, the interpretation here is rather
different --- remember, these are the $\sigma(a)$ and $\vth(a)$
required to have $V(a)\equiv 0$; and so a transition layer that is
dynamically stable (to radial perturbations) at all $a$.

%------------------------------------------------------------
%\subsubsection{Mazur--Mottola limit}
%------------------------------------------------------------

In contrast the Mazur--Mottola model corresponds to another particular
limiting choice for the energy density in the de Sitter vacuum:
\begin{equation}
k (2M)^3 = M \qquad \implies \qquad k = {1\over8M^2}.
\end{equation}
To understand the nature of this limit it is convenient to write
\begin{equation}
k ={1\over8M^2(1+\epsilon)^2}.
\end{equation}
with $\epsilon \gsim0$. Then the energy density and surface tension
in (\ref{E:sigma-0}) and (\ref{E:theta-0}) are both real for
\begin{equation}
a \in (2M, 2M[1+\epsilon] ).
\end{equation}
This indicates, purely on kinematic grounds, a severely restricted
range of possible motions for the shell.

In the limit that $\epsilon\to0$ the surface energy density of our
3-layer model is forced to go to zero. In the language of the 5-layer
Mazur--Mottola model~\cite{gravastar} this means
\begin{equation}
\sigma_{\mathrm{net}} = 
\sigma_+ 
+ 
\rho_{\mathrm{transition}} \; \eta_{\mathrm{transition}} 
+ \sigma_- \to 0.
\end{equation}
In other words, as the transition layer in the Mazur--Mottola 5-layer
model becomes thinner and thinner, the contributions from $\sigma_\pm$
conspire with the density of the stiff matter in the transition layer
to produce a net surface energy density of zero. A quick way of seeing
that this \emph{must} be the case is to realise that if one is
attributing all of the mass of the black hole to the de Sitter vacuum
in the interior, then whatever is going on in the transition layer
must average out to zero surface energy density.

In contrast, the transverse pressures in the transition layer do not
have any pleasantly behaved limit. Indeed $\vth\to-\infty$,
corresponding to infinite transverse pressures. This is not
unexpected, and is one of the reasons many in the general relativity
community are extremely hesitant to adopt models with thin layers of
matter hovering just at the location where the event horizon would
have been expected to form.

%------------------------------------------------------------
\subsection{Stiff-shell gravastars}
%------------------------------------------------------------

Given the potential problems associated with the limits just
discussed, there is another model among the class we have considered
which might be considered to be an alternative 3-layer simplification
of the Mazur--Mottola model. In particular, let us specify the
equation of state of the thin shell to be that of stiff matter,
$p=\rho$ or $\vth=-\sigma$.  This is a limit of the Mazur--Mottola
model in which the equation of state of the thick stiff matter shell
is maintained, but the shell itself is made infinitesimally thin. One
principle difference is that no attempt is made to equate the energy
of the interior de Sitter vacuum with the ADM mass of the exterior
solution. The other crucial point is that the thin shell is to be
placed at $a>2M$, where one avoids infinite transverse pressures
discussed above.

The analysis of the conditions required for satisfying the dominant
energy condition in section \ref{secV} have already shown that such
solutions exist and are dynamically stable to spherically symmetric
perturbations. This follows since the stiff matter shell occurs at the
extreme values for which the dominant energy condition holds.  In
particular, we found that such solutions exist if $kM^2$ is less than
or equal to the critical value given by (\ref{lambda_cr}).
Furthermore, for $\la<\la\ns{cr}$ there are two values of $a$ at which
we can place a stiff shell in stable gravastars, the lower value, $a_1$,
being in the range $2M<a_1<2.30056\, M$. For the case plotted in
figure \ref{la72}, for example, $a_1=2.12432\,M$ and $a_2=3M$. The
inner stiff shell case is certainly so close to the putative horizon
that any test to distinguish such an object from a true Schwarzschild
black hole would be extremely difficult in astrophysical contexts.

%------------------------------------------------------------
\subsection{Anti-de Sitter gravastars}
%------------------------------------------------------------

Although we have been primarily interested in the case of a de Sitter
interior, our general analysis remains valid for the case of an
anti-de Sitter interior, with $\Lambda<0$. For such interiors it is
impossible to satisfy the junction conditions with a thin shell having
a stiff equation of state. However, it is possible to take a thin
shell with a ``cosmological constant'' or ``vacuum energy'' equation
of state, $p=-\rho$, or $\vth=+\sigma$.

Since $k<0$, there are a number of changes to the analysis of section
\ref{secV}. In particular, there is now no upper bound to the value of $a$.
The surface energy, $\sigma$, is always positive, being bounded below
by (\ref{sigma_bound}). The strong energy condition is now violated
for particular parameter values. We are interested in the particular
case $\vth$=$\sigma$. Applying this condition to (\ref{E:sigma-V}) and
(\ref{E:theta-V}) we have
\begin{equation}
\left(1-{3M\over a}\right)\sqrt{1-2ka^2}=\sqrt{1-{2M\over a}}.
\label{anti-stiff-1}
\end{equation}
Squaring (\ref{anti-stiff-1}) and rearranging we find that the shell
must be located at values of $a$ which solve the quartic
\begin{equation}
2{a}^{4}k-12{a}^{3}Mk+18M^2ka^2+4Ma-9M^2=0\,,
\label{anti-stiff-2}
\end{equation}
and lie in the range $a>3M$. An analysis of the discriminant of this
quartic reveals that solutions exist for all values $k<0$.
Furthermore, for each value of $k<0$ there is only one root of
(\ref{anti-stiff-1}) in the range $a>3M$.  For small values of $-kM^2$
one finds that that $a$ is very large, but as $-kM^2$ gets larger the
value of $a$ approaches $3M$ asymptotically.  For $kM^2=-1$, for
example, $a=3.44898\, M$.

%------------------------------------------------------------
\section{Discussion}
%------------------------------------------------------------

The general relativity community is, by and large, extremely
comfortable with the ``black hole'' concept; a concept firmly based in
the classical solutions of the Einstein
equations~\cite{membrane,wald,schutz}. In contrast, there is a strong
undercurrent [certainly not the mainstream] in the particle physics
and condensed matter communities that views the notion of event
horizons with some alarm --- this has lead over the years to repeated
suggestions that quantum physics should intervene at/ near the the
would-be event horizon, either replacing it with a singularity or
preventing appearance of the horizon in the first place. A particular
proposal along these lines that has recently attracted considerable
attention is the Mazur--Mottola ``gravastar''~\cite{gravastar}.

In this article we have considered the dynamical stability of a
3-layer simplification of the 5-layer Mazur--Mottola model.
Aficionados of the model will be happy to know that there are many
equations of state for the transition layer that imply dynamical
stability for the gravastar configuration. Those who are less
enamoured of the model will be equally happy to see that large classes
of equation of state are ruled out. Our own interpretation of these
results is that this calculation, or suitable modification thereof, is
a necessary first step towards \emph{any} kind of serious model
building with a thin transition layer separating two ``vacuum''
regions.

In this regard, we have in this article explored a tentative
suggestion: If one really wishes to build a 3-layer thin-shell
alternative to a black hole, then it seems to us that $r=2M$ is not
the appropriate place to insert the transition layer. In order to
avoid the infinite stress (presumably cutoff at the Planck scale so
that we would really be talking about Planck-scale stresses) that
occur when the transition layer is placed at $r=2M$, it would seem to
us to be more profitable to move the transition layer out to $r>2M$.

We have demonstrated that particular models do exist which realise
this possibility.  One particular case is the stiff shell gravastar,
namely an exterior Schwarzschild geometry, and an interior de Sitter
vacuum, separated by a stiff matter thin shell. Such models could be
regarded as simplified versions of the Mazur--Mottola model which
avoid its singular limit.  Stiff shell gravastars exist and are
dynamically stable for $\Lambda\le 6\la\ns{cr}/M^2=0.145827\,/M^2$.
In fact, generally two possible stiff shell gravastars exist, and in
one of these cases one must place the thin shell at a value $2M<r<
2.30056\,M$. Since this is very close to the nominal position of a
black hole horizon, it would be extremely difficult to distinguish the
exterior geometry of such a gravastar from a genuine black hole in
astrophysical contexts.

The idea of moving the transition layer to a finite $r>2M$ should
possibly also be explored in the related
Chapline--Hohlfeld--Laughlin--Santiago
model~\cite{laughlin,laughlin2,laughlin3}. More generally, if the
transition layer is not ``thin'' but rather of finite thickness, then
investigations along the lines of Gliner's~\cite{gliner} and
Dymnikova's~\cite{dymnikova} ideas might be profitable.

Finally we emphasise what we feel is the most important technical
aspect of this paper. We have analyzed spherically symmetric thin
shells in a large class of background geometries [arbitrary $m(r)$],
and related the resulting shell motion to an equivalent
``non-relativistic particle''. Furthermore we have shown how to take
any desired ``potential'' $V(a)$ and ``invert'' it to determine the
equation of state for shell-matter that would lead to this potential.
This result is of general interest to anyone building thin-shell
models --- regardless of one's views on gravastars versus black holes.
\bigskip

\noindent {\em Note added:} Since this paper was completed we have
learned that solutions with features similar to those discussed here
-- possessing an asymptotically Schwarzschild exterior, a thin-shell
and a de Sitter core -- have recently been found by Wohlfarth
\cite{Wohlfarth} in a gravity model in which the Einstein--Hilbert action
is replaced by a ``Born-Infeld style'' action nonlinear in the curvature.
Wohlfarth's solutions completely regularize the Schwarzschild singularity
for $M>0$. While much work remains to be done, these results suggest
that the model presented in this paper may represent a simplified
version of a class of objects which occur very naturally in string
theory inspired gravity models.
%---------------------------------------------------
\appendix
%---------------------------------------------------
 
%------------------------------------------------
\ack
%------------------------------------------------
    
We would like to thank Benedict Carter for his careful reading and
comments on an early version of the manuscript.  This research was
supported by the Marsden Fund administered by the Royal Society of New
Zealand.

%------------------------------------------------
\section*{References}    
%-----------------------------------------------------------------

%-----------------------------------------------------------------
\end{document}